\documentclass[conference]{IEEEtran}

\IEEEoverridecommandlockouts
\usepackage{cite}
\usepackage[frozencache,cachedir=.]{minted}

\usepackage{amsmath}
\usepackage{minted}
\usepackage[caption=false]{subfig}
\usepackage{multirow}
\usepackage[usenames,dvipsnames]{color}
\usepackage{algorithm}
\usepackage{tikz} 
\usepackage{algorithmic}
\usepackage{svg}

\usepackage{amsmath,amssymb,amsfonts}
\usepackage{graphicx}
\usepackage{algorithmic}
\usepackage{algorithm}
\begin{document}



\def\x{{\mathbf x}}
\def\L{{\cal L}}

\title{Adaptive Services Function Chain Orchestration For Digital Health Twin Use Cases: Heuristic-boosted Q-Learning Approach  \\
{\footnotesize}
\thanks{The EPI project is funded by the Dutch Science Foundation in the Commit2Data program.}
}

\author{\IEEEauthorblockN{1\textsuperscript{st} Jamila Alsayed Kassem}
\IEEEauthorblockA{\textit{MNS Lab, IvI} \\
\textit{University of Amsterdam}\\
Amsterdam, Netherlands \\
j.alsayedkassem@uva.nl}
\and
\IEEEauthorblockN{2\textsuperscript{nd} Li Zhong}
\IEEEauthorblockA{\textit{SNE, IvI} \\
\textit{University of Amsterdam}\\
Amsterdam, Netherlands\\
li.zhong@student.uva.nl}
\and
\IEEEauthorblockN{3\textsuperscript{rd} Arie Taal}
\IEEEauthorblockA{\textit{MNS, IvI} \\
\textit{University of Amsterdam}\\
Amsterdam, Netherlands \\
a.taal9@upcmail.nl}
\and
\IEEEauthorblockN{4\textsuperscript{th} Paola Grosso }
\IEEEauthorblockA{\textit{MNS, IvI} \\
\textit{University of Amsterdam}\\
Amsterdam, Netherlands \\
p.grosso@uva.nl}

}

\maketitle

\begin{abstract}
Digital Twin (DT) is a prominent technology to utilise and deploy within the healthcare sector. Yet, the main challenges facing such applications are: Strict health data-sharing policies, high-performance network requirements, and possible infrastructure resource limitations. In this paper, we address all the challenges by provisioning adaptive Virtual Network Functions (VNFs) to enforce security policies associated with different data-sharing scenarios. 

 We define a Cloud-Native Network orchestrator on top of a multi-node cluster mesh infrastructure for flexible and dynamic container scheduling. The proposed framework considers the intended data-sharing use case, the policies associated, and infrastructure configurations, then provision Service Function Chaining (SFC) and provides routing configurations accordingly with little to no human intervention.

Moreover, what is \textit{optimal} when deploying SFC is dependent on the use case itself, and we tune the hyperparameters to prioritise resource utilisation or latency in an effort to comply with the performance requirements. As a result, we provide an adaptive network orchestration for digital health twin use cases, that is policy-aware, requirements-aware, and resource-aware. 
\end{abstract}


\section{Digital Health Twin: The concept}
\label{sec:intro}

The concept of Digital Twins (DTs) is not new, and it originally appeared in the early 1990s \cite{gelernter_1991} under different terminologies, such as the "Mirror Space Model" \cite{grieves_2006}, "Information Mirror Mode"\cite{grieves_2005}, etc. Although DT has been primarily discussed within engineering and industrial contexts, medical and health use cases are not excluded from the DTs' impact. In practice, a DT is a digitalised model that dynamically couples both virtual and physical twins, and makes use of contemporary technologies like smart sensors (IoT devices) and data analytics in order to predict and identify failures, discover and simulate optimising opportunities, and improve outcomes \cite{kamel}.

Deploying a Digital Health Twin (DHT) utilises medical data-sharing and patient-generated data to empower personalised medicine. With that in mind, DHTs come as a natural, complementary approach to implementing personalised medicine, since it offers the capacity to model a distinct patient. In the EPI project\footnote{https://enablingpersonalizedinterventions.nl/}, we develop a framework to combine data analytics, and health decision support algorithms to create personalised insights for prevention, management, and intervention to providers and patients. We start by defining a number of data-sharing use cases to effectively utilise patients-generated data within the EPI consortium:

\begin{itemize}
    \item \textbf{EHR repository use case:} We need to be able to build an Electronic Health Records (EHR) repository (centralised or decentralised) and have remote access to a patient's medical history, where this comes as a natural first step to start building a DHT model.
    \item \textbf{ML model sharing use case:} With this use case we are supporting another feature, such that we need to be able to process medical data by applying aggregation, analytics, and Machine Learning (ML) algorithms to have predictive and informed responses to a physical's system status. 
    \item \textbf{Heathcare data streaming use case:} We need to have a real-time status update of the patients for monitoring, control, and data acquisition. 
\end{itemize}

The adoption of DHTs is accompanied and accelerated by maturing and growing computing technologies. This is led by cloud computing and network virtualisation, which offers the means to facilitate knowledge discovery by provisioning on-demand computing and network resources. It is evident that one of the main contributing factors toward the successful and reliable deployment of DHTs is the underlying network paradigm connecting all the data-sharing components to effectively run a use case \cite{mashaly}. It is crucial for the network to abide by a set of requirements set while running a use case: policy-wise, latency-wise, and resource-wise.

\section{Health Data sharing Policies}
\label{policy}


In previous work \cite{9211394}, we defined the collaboration logic model which the EPI data-sharing framework follows to aggregate higher-level data-sharing agreements, with lower-level network security goals to establish a policy-abiding data-sharing session. This can be further translated and aggregated to be enforced at a lower level. The policy additionally depends on the parties involved, and the data type being shared. After discussions with the hospitals within the EPI consortium, we can enumerate the following security goals as: 1) Providing access control to the data resources, 2) Identifying and authenticating parties, 3) Health data integrity, confidentiality,  and 4) non-repudiation.

These security goals can be achieved by applying different security mechanisms, such as deploying access control and security protocols namely, SSL, SSH, IPSec, firewalling, and the security gateway systems \cite{laborde_kamel_barre_benzekri_2007}. We take a separation of concerns approach where we define two levels of policies: data level, and network level. We assume that all policies fall under one of the two levels, and this is further discussed in \cite{9973688}.

By virtualising the network services, we aim to deploy and provision on-the-fly exemplary reliable Virtual Network Functions (VNF), which we call Bridging Functions (BFs), that can accomplish these security goals. The framework's goal is to define an adaptive BFs Chain (BFC) orchestrator, enforcing all types of network policies. We provide an access control mechanism by containerising a ready-to-deploy firewall function, and we address the rest by implementing standard security protocols to encrypt traffic. Once the security goal is specified, then we can map that to the mandated network services, and to the defined enforcement primitives: \textit{Filter} traffic (\textbf{F}) and/or \textit{transform} traffic (\textbf{T}).

The framework dynamically provisions these services by placing the BFs on available N-PoPs (Network Points of Placements), assigning the service requests to the running function, and routing traffic along the function's chain to enforce a policy. Along with the data-sharing policy changes, the available N-PoPs are constrained with different use cases running. As a result, the high-level policy is defined and specified by two 3-tuples: $<actors, acts, inRelation>$, and $<endNodes, BFC, N\text{-}PoPs>$ , where the second one is in accordance to the lower-level network policy. 

\section{DHT use cases and N-PoP restrictions}
\label{requirements}


The three use cases, as previously defined,  describe the requirements, restrictions, and network configurations associated with each one.

\subsection{Electronic Health Records Repository}
\begin{figure}[!ht]
    \centering
    \includegraphics[width=\columnwidth,trim={0 3cm 0 0cm},clip]{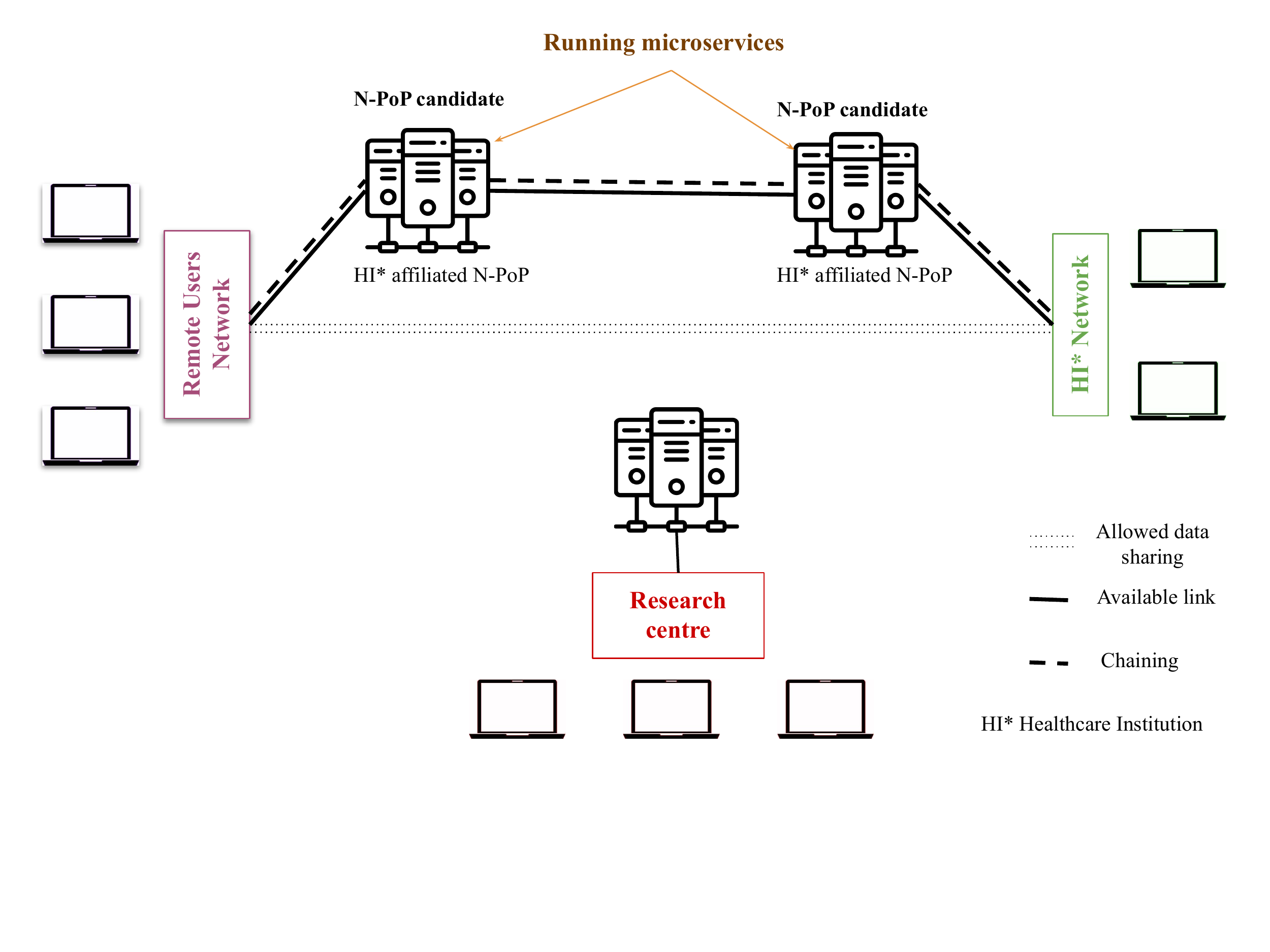}
    \caption{The infrastructure graph configuration under the EHR and the Streaming use cases.}
    \label{fig:ehr}
\end{figure}

This use case is built to run EHR (Electronic Health Records) data-sharing scenarios where there are two N-PoPs affiliated with healthcare institutions (HI), and an isolated third-party research centre N-PoP. Fig. \ref{fig:ehr} illustrates the use cases' configuration and the infrastructure setup, where the data-sharing movement is expected from a remote user network to HI  network  and vice versa. This use case requires remote access to sets and effective queries (Update/Insert/Get) of a patient’s medical records. 

The third-party research centre is uninvolved in this transaction, and hence the affiliated third N-PoP is isolated from the rest. The placement algorithm will not consider it while placing the BFCs request and will place the functions on the other two to secure network traffic according to the BFC request.  


\subsection{Machine Learning model sharing}
In an effort to advance healthcare research, we need to accelerate and support the deployment of ML-featured applications (such as psychiatry diagnosis, effective drug prescriptions, and side effects predictions, etc.), we define this use case where ML and analytics algorithms are sent from a research center to be trained on data residing in the HI. Moreover, data movement is allowed again from the data provider (the HI) back to the algorithm provider (the research centre), so that the distributively trained model can be joined back again into one (more accurate) model, and shipped back to be ready to use. 

\begin{figure}[!ht]
    \centering
    \includegraphics[width=\columnwidth ,trim={0 3cm 0 0cm},clip]{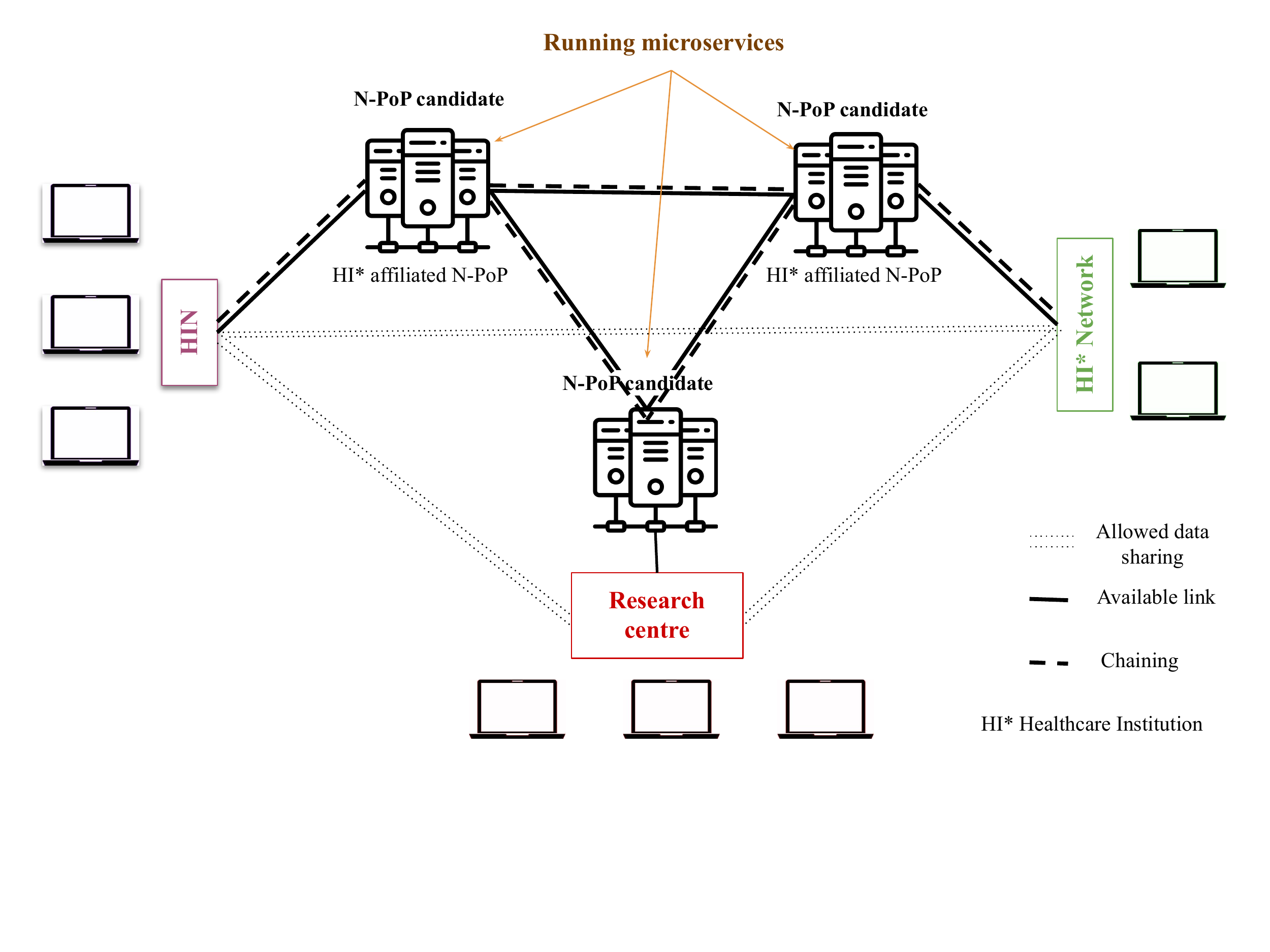}
    \caption{The infrastructure graph configuration under the ML model sharing use case.}
    \label{fig:ml}
\end{figure}

As a result, the policy dictates the availability of links across affiliated N-PoPs to ensure who and where data is handled, as illustrated in Fig. \ref{fig:ml}. After establishing those restrictions on traffic flow, we then reconfigure the network to adapt to that, and secure the network traffic. 

\subsection{Healthcare data streaming}
This use case describes health data streaming, an example application is to monitor the patient's status via wearable data, provide timely interventions, etc. The data involved in this use case is sensitive data, which means that similarly as in use case A, HI-affiliated N-PoPs are only considered for placement.


\section{BFC provisioning}
\label{provision_model}


\subsection{Adaptive Provisioning of BFC}
We need to adapt the network to performance and policy requirements by automating the provisioning of the network service function chains. To do that, we consider multiple approaches to provide a best-effort placement of network microservices on a Kubernetes cluster mesh of N-PoPs middleboxes. We consider the following provisioning steps: 
\begin{enumerate}
    \item Profiling of BFC service chains: Determine the computing and network profiles while running microservices within the service chain under different use cases.
    \item Mapping BFC requests to running microservices: Assign new requests to run microservices' deployments under different constraints. 
    \item Allocation of N-PoP: When mapping fails, place a new instance on an N-PoP. This is also done according to a set of placement rules/constraints.
    
    \item Chaining the microservices: This step is to assign available (routable) links and host virtual links and actually chain the microservices according to internal packet flows requirements between the microservices pairs.
\end{enumerate}


\subsection{Provisioning Decisions model}
\subsubsection{Infrastructure}
The currently available infrastructure is represented as a directed, attributed graph $G(C, L, CPU^C, D^L)$. The finite set of vertices $C$ represents the clusters that serve as network points of placements (N-PoP), where a network microservice can be placed. The unidirectional link between clusters is represented by the set  $L \subseteq C \times C $, where $(i,j) \in L$ represents the unidirectional link between cluster $c_i$ and $c_j \in C$.  

Within the infrastructure environment, there exist limited computing and network resources. These are represented by the attribute sets $CPU^C$ and $D^L$. ${cpu^C}_i \in CPU^C$ is the maximum CPU capacity of cluster $c_i \in C$, and ${d^L}_{(i,j)} \in D^L$ is the network capacity with an associated delay of link $(i,j) \in L$ at time $t$. 

The configuration of $G$ is representative of the data-sharing policy held by the infrastructure providers within the EPI consortium, such that for any cluster $c_i$ there is a cluster $c_j$ considered as an N-PoP candidate under the condition that:
\begin{equation}
    (i,j) = 
    \begin{cases}
    1 | exists, & \ \text{$\iff c_j$ is considered N-PoP candidate} \\
    0, & \ otherwise.
    \end{cases}
    \label{eq:1}
\end{equation}
The lack of an edge would drive the placement algorithm to either centralise placement on $c_i$ or distribute it on a different cluster $c_z$ such that $(i,z)$ exists.

As an example, when data provider A and algorithm provider B are sharing data, the policy dictates that a third-party associated N-PoP should not be considered for placement. 

\subsubsection{BFC requests and profiles}
$BFC$s are ordered sets of network service requests, and a single request can be composed of multiple chained microservices. To successfully provision these requests, microservices can be placed and hosted on one or more clusters. $BFC=\{f_1, ..., f_n\}$ is a first-come-first-placed ordered set, where to each request $f \in BFC$ a directed graph $G_f=(S_f, L_f)$ is associated. Service requests are allowed to run as long as required, and multiple service requests can populate the infrastructure at any time. The finite set $S_f$ represents the microservice functions that need to be assigned to run a request $f \in BFC$. 

Moreover, $S_f \subseteq \mu S$, the list of all possible containerised network microservices (\textit{e.g.} firewall, encryption, load balancer, NAT, etc.). The edges of the graph are members of the set $L_f \subseteq S_f \times S_f $ and are associated with inter-virtual links between two microservices, where $(q,r) \in L_f$ represents the unidirectional link between requested microservices ${\mu s}_q$ and ${\mu s}_r \in S_f$.

Members of $\mu S$ can be instantiated multiple times and can be deployed as a microservice replica with different configurations, such that ${\mu s_m}^n$ is the $n^{th}$ instance of microservice $\mu s_m \in \mu S$. With that, one can instantiate smaller and bigger instances of the same microservice. 

To model this, we define $CPU\{{\mu s_m}^n\}$ and $d\{{\mu s_m}^n\}$, the computing capacity and processing delay of the $n^{th}$ instance of microservice $\mu s_m$. On the other hand, placed and running microservices can be shared across multiple incoming network service requests, as an example, ${\mu s_m}^n$ can be active for $f_1$  and $f_2$. 


A network service request has profiled requirements values that may differ under different use cases utilising the requested service chain setup. As an example, a service request $f$ active for a streaming use case needs more CPU resources and higher bandwidth than active for another use case. To model this, we define $CPU_{q,f}^{u}$, $D_{f}^{u}$
, representing the required CPU running microservice ${\mu s}_q$ when active for $f$ under use case $u \in U$, the end-to-end maximum delay respectively. ($U$ is the set of all possible use cases)



\subsection{Variables}
The goal of the proposed placement algorithm is to allocate network microservice instances to N-PoP clusters and then map the deployed microservices to the running instance. Moreover, the algorithm considers the use cases being deployed over the network infrastructure, and tailors the placement according to profiled resource consumption per use case. Furthermore, the microservices need to be chained to enforce the network policy route.

The first variable we consider is the virtual link mapping function $M_{(i,j),(q,r),f}  ^C$, where we decide to utilise link $(i,j)$ with virtual link $(q,r)$ under request $f$ such that:
\begin{equation}
    \resizebox{\columnwidth}{!}
     {$
    M_{(i,j),(q,r),f}  ^C  = 
    \begin{cases}
    1, & \ \text{link $(i,j) \in L$ is mapped to link $(q,r) \in L_f$} \\
    0, & \ \text{otherwise.}
    \end{cases}
    \label{eq:2}$}
\end{equation}

Then we consider is the placement function $P_{c_i, \mu s_m^n}^C \in \{0,1\}$, such that:
\begin{equation}
    P_{c_i,{\mu s_m}^n}^C  = 
    \begin{cases}
    1, & \ \text{${\mu s_m}^n$ is placed on $c_i \in C$} \\
    0, & \ otherwise.
    \end{cases}
    \label{eq:3}
\end{equation}
We also consider the mapping function $M_{\mu s_m^n,q,f} \in \{0,1\}$ to assign a microservice ${\mu s}_m \in \mu S$ needed by the service request $f$, and associated with ${\mu s}_q \in S_f$, to an instance ${\mu s_m}^n$ running on the cluster $c_i$, such that:
\begin{equation}
    M_{\mu s_m^n,q,f} = 
    \begin{cases}
    1, & \ \text{microservice ${\mu s_m}^n$ is mapped to $\mu s_q$} \\
    0, & \ otherwise.
    \end{cases}
    \label{eq:4}
\end{equation}

Note that variables defined in \textit{Eq.} \ref{eq:2}, \ref{eq:3}, \ref{eq:4} are dependant values such that, suppose that we have a subgraph of $S_f$: \\
\begin{center}
\begin{tikzpicture}[node distance={20mm}, thick, main/.style = {draw, circle}]
\node[main] (1) {$\mu s_q$}; 
\node[main] (2) [right of=1] {$\mu s_r$}; 
\draw[->] (1) -- (2);
\end{tikzpicture}
\end{center}
In fact the possibility to connect microservices depends on their mapping and placement. For example, we can have a successful microservice provisioning of $\mu s_q$ can be that: $M_{\mu s_m^n,q,f} =1$ and $P_{c_i,\mu s_m^n}^C  =1$. While the output of provisioning $\mu s_r$ such that $M_{\mu s_k^p,r,f} =1$, and $P_{c_j,\mu s_k^p}^C  =1$ is conditional to the existence of the link $(i,j)$. This means that virtual link mapping  $M_{(i,j),(q,r),f}  ^C= 1$, and it is determined, such that: 

\begin{equation}
    \resizebox{\columnwidth}{!}
     {$
    M_{(i,j),(q,r),f}  ^C  =
    1 \implies \text{$\exists M_{\mu s_m^n,q,f} = 1 , P_{c_i, \mu s_m^n}^C =1 \land \exists M_{\mu s_p^k, r,f}=1  P_{c_j, \mu s_p^k}^C =1 $} \\
$}
\end{equation}
Before assigning a microservice of flow request $f$ to cluster $c_i$, we define the available CPU resources of $c_i$ as:

\begin{equation}
      cpu_i^C - \sum_{\forall \mu s_m^n | {\mu s_m}\in \mu S} P_{c_i,\mu s_m^n}^C  . CPU\{\mu s_m^n\}
\end{equation}

The expected latency on utilised links running flow request $f$ at time $t$ is:
\begin{equation}
   \hat{ D}_{L,f} = \sum_{\forall (i,j) \in L, (q,r) \in L_f} M_{(i,j),(q,r),f}^C . d_{(i,j)}^L ,
\end{equation}
\begin{equation}
   \hat{D}_{C,f} = \sum_{\substack{\forall c_i \in C, \\  \mu s_m^n | {\mu s_m} \in \mu S, \\ {\mu s_q} \in S_f}} M_{\mu s_m^n,q,f} . d\{\mu s_m^n\}
    \end{equation}


The un-utilised resources on already running microservice instances are approximated to be:
\begin{equation}
    \resizebox{0.9\columnwidth}{!}
     {$cpu_i^C - \sum\limits_{f \in BFC} \sum\limits_{\substack{\forall c_i \in C, \\ \mu s_m^n | \mu s_m \in \mu S, \\ \mu s_q\in S_f}}  M_{\mu s_m^n,q,f}  . \sum\limits_{u \in U} CPU_{q,f}^{u}$}
\end{equation}

\subsection{Objectives and placement constraints}
The objective is to minmise end-to-end latency and maximise CPU utilisation across the infrastructure's clusters, such that:
\begin{equation}
    min \sum_{\substack{\forall c_i \in C, \\ \mu s_m^n |  {\mu s_m} \in \mu S}}  P_{c_i,\mu s_m^n}^C ,
\end{equation}

\begin{equation}
    min (\hat{D}_{C,f} + \hat{D}_{L, f}) \leq D^u_f,
\end{equation}
The two minima might not always correlate depending on the use case, hence the provisioning tools should prioritise minimising one over the other when appropriate. 
\subsection{Possible constraints} 
The first constraint is to ensure that the allocated function instances on the cluster $c_i$ do not exceed available CPU resources, such that:
\begin{equation}
   \forall c_i: \sum_{\forall \mu s_m^n | {\mu s_m} \in \mu S} P_{c_i,\mu s_m^n}^C . CPU\{\mu s_m^n\} \leq CPU_i^C
\end{equation}

The second constraint is to ensure that the mapped microservice requested does not exceed the available CPU at the running instance:

\begin{equation}
\resizebox{\columnwidth}{!}{
  $ \sum\limits_{\substack{{\mu s}_ q \in S_f,\\ f \in BFC}} M_{\mu s_m^n,q,f}\sum\limits_{u \in U} CPU_{q,f}^u \leq  CPU\{\mu s_m^n\} $}
  \label{eq:c2}
\end{equation}
Eq. \ref{eq:c3} is to ensure that the maximal latency allowed is greater than the delay composed of link and processing delays.

\begin{equation}
\resizebox{\columnwidth}{!}{

     $\forall u: \sum\limits_{\substack{\forall \mu s_m^n | {\mu s_m} \in \mu S, \\  f \in BFC, \\ \mu s_q \in S_f}} M_{\mu s_m^n,q,f} . D\{{\mu s_m}^n\} + \\ \sum\limits_{\substack{\forall (i,j) \in L, \\ f \in BFC, \\ (q,r) \in L_f}} M_{(i,j),(q,r),f}  ^C  . D_{(i,j)}^L \leq D_{f}^u $}
  \label{eq:c3}
\end{equation}

If a cluster $c_i$ is chosen for placement of $n^{th}$ microservice of ${\mu s_m}$ requested by $f$, the function needs to be running (placed) before assignment.
\begin{equation}
    M_{\mu s_m^n,q,f}   \leq P_{c_i,\mu s_m^n}^C  
\end{equation}
All microservices requested by request $f$ are placed and mapped to the infrastructure.
\begin{equation}
    \forall f \in BFC: \sum_{\substack{\forall \mu s_m^n | \mu s_m \in \mu S, \\ \mu s_q \in S_f}} M_{\mu s_m^n,q,f}=1
\end{equation}
Building virtual paths over available links must obey the following rule to ensure that there exists a mapped link $(i, j)$  to $(q, r)$ if ${\mu s}_q$ is placed on $c_i$ and ${\mu s}_r$ is placed on $c_j$, such that $\mu s_m^{n\prime}$ is potentially a different microservice instance assigned to $\mu s_r$:

\begin{multline}
    \forall \mu s_q, \mu s_r \in S_f, (q,r) \in L_f: \\ M_{\mu s_m^n,q,f} . P_{c_i,\mu s_m^n}^C. M_{\mu s_{m\prime}^{n\prime},r,f}^C . P_{c_j,\mu s_{m\prime}^{n\prime}}  = M_{(i,j),(q,r),f}^C 
\end{multline}
Lastly, the microservice should not run indefinitely once instantiated, but the microservice is deleted after being idle for a specified duration of time $T$, such that $P_{c_i,\mu s_m^n}^C=0 \iff \sum^{t=t'+T}_{t=t'}CPU\{\mu s_m^n  \}$.
\section{Provisioning Approaches}
\begin{figure}[ht!]
\centering
\includegraphics[width=\columnwidth, height=5cm]{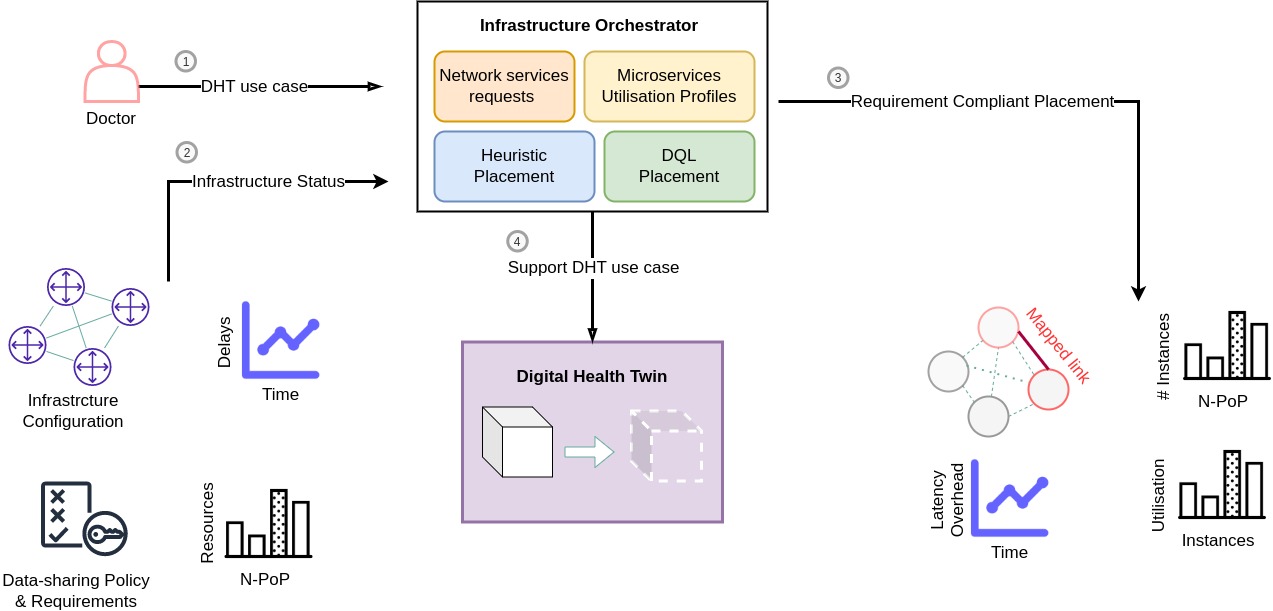}
\caption{A high-level overview of the orchestrator.}
\label{fig:arch}
\end{figure}

We deploy three approaches in an effort to satisfy the use cases' requirements: a greedy heuristic approach, Deep Q-Learning (DQL), and a Heuristic-boosted DQL (HDQL). Fig. \ref{fig:arch} illustrates the high-level overview of the infrastructure orchestrator with the different components to make the provisioning decisions.

We first deploy a greedy-based  heuristic approach to reduce complexity with increasingly complex network policies and use cases, and still provide a manageable best-effort provisioning decision by choosing the first N-PoP candidate meeting the placement constraints. This approach is dependent on accurate CPU profiles and does not react to resource usage and network latency bursts and anomalies. It is not concerned with the most optimal provisioning choice, instead, it provides the decisions that work.

\subsection{Greedy Heuristic BFC Deployment Algorithm}
\begin{algorithm}[ht!]
\caption{Heuristic BFC Deployment Algorithm}\label{alg:heuristic}

\begin{algorithmic}[1]
\REQUIRE $G(C,L)$, $G_f(S_f, L_f)$
\ENSURE $P_{c_i,\mu s_m^n}^C, M_{\mu s_m^n,q,f}$ 

\WHILE{Not all $\mu s_q$ has been mapped to $c_i \in C$}
\WHILE{Not all $c_i \in C$ has been checked}
\IF{$c_i \in$ N-PoP candidates}
\IF{$\mu s_m^n$ meets requirements}
\STATE $M_{\mu s_m^n, q,f} = 1$
\ENDIF
\ENDIF 
\ENDWHILE
\ENDWHILE
\IF{Not all $\mu s_q$ mapped to C}
\WHILE{Not all $c_i \in C$ has been checked}
\IF{${\mu s}_m^{n'}$ meets requirements}
\STATE $P_{c_i,\mu s_m^{n'}}^C \wedge M_{\mu s_m^{n'}, q,f} = 1 $
\ENDIF
\ENDWHILE
\ENDIF
\label{algo:heuristic}
\end{algorithmic}
\end{algorithm}

This algorithm loops over the requested microservices and N-PoP candidates (lines 1-2), if a running microservice instance $\mu s_m^n$ meets the requirements set according to placement properties and constraints in the previous section (line 4) then, $\mu s_q$ is assigned to run on the cluster $c_i$ and the loop continues to consider more microservice requests. If the assignment of a microservice failed on all clusters, then new instances need to be placed, and the algorithm loops again over clusters, but this time places new instances and then assigns them to a microservice request. The algorithm applies constraints defined in Eq. (12), (13), (14), (15), and (16) in lines 4 and 12, and it does so according to CPU and delays profiles. We only instantiate and place a new service when all the mapping fails (in line 13), and by that we prioritise minimising placement, as in Eq. (10) under the constraints of delay.

\subsection{DQL Algorithm}
The performance of the first approach is highly dependant on the accuracy of the CPU and delays profiles. Compared to traditional heuristic-based resource scaling methods, Reinforcement Learning-based (RL) solutions are equipped to deal un-profiled network and resource bursts, instead this DQL-approach relies on querying the current state and reacting via provisioning actions to maximise performance rewards. 


\textbf{Action space:}
The discrete actions of microservices placing and assignment are structured as $[$Cluster ID, Place/ Map/ Destroy, Instance ID, Microservice ID, Proxy ID$]$. The first value specifies the cluster that is considered; the N-PoP candidate. The second value specifies the type of provisioning action: placing a new instance, mapping a running instance to a request, or deleting an idle microservice instance. The third and fourth values, respectively, refer to the microservice instance and the type of microservice ($n$ and $m$ within $\mu s^m_n$). Lastly, the Proxy ID identifies the proxy we are configuring to chain the instantiated microservices and handles the network services requests. \\ 

\textbf{State Space:}
The state space consists of the infrastructure's variables, which the DQL interacts with to monitor the environment's changes based on decisions. The provisioning is done across multiple sites, and the DQL algorithm needs to make a satisfactory trade-off between resource cost and latency. Therefore, the state of the agent should contain information about: CPU utilisation/cluster, placement status of microservices, the number of microservices across the clusters, and the response time of requests. \\  


\textbf{Reward Function:}
The reward function measures the performance incentive for the agent to perform a new action, based on the infrastructure's current state. The entire reward $R_{all}$ at time t is calculated as the weighted sum of reward in resource cost $R_{res}$ and reward in performance $R_{perf}$, which is shown in:
\begin{equation}
    R_{all} = \alpha R_{res} + \beta R_{perf}
    \label{eq:reward}
\end{equation} 
Hyperparameters $\alpha$ and $\beta$ are used to control the importance of these two values compared to the entire reward. One effective definition of reward function steers the agent towards better performance with higher utilisation of resources (prioristing Eq. (10) vs Eq. (11)).




\textbf{Action Policy:}
In the traditional DQL approach, the policy is set to map an observable state $s_t$ to a provisioning action $a_t$ at a time $t$. The policy is optimised by learning the Q-value performing  $a_t$ in state $s_t$, according to the following formula:
\begin{equation}
\label{policy-rl}
    \pi(s_t) = 
    \begin{cases}
    max_{a_t}Q(s_t,a_t),  & \ if q \leq p \\
    a_{random}, & \ otherwise
    \end{cases}
\end{equation}
, where $q$ is a random value  with uniform probability in $[0,1]$, and $0 \leq p \leq 1 $ is the exploration/exploitation ratio parameter.

\subsection{Heuristic-boosted Algorithm}

While the previous tool can reactively adapt the network's configuration to optimally provision BFC requests, it still can take a long time to converge with increasingly complex requests, use cases, and N-PoP configurations. Subsequently, we propose to combine both provision approaches (A and B) to deploy a Heuristic-boosted DQL provisioning of SFC. We aim to accelerate the decision-making, and guide the model learning via a new action policy:
\begin{equation}
        \resizebox{\columnwidth}{!}{
    $\pi(s_t) = 
    \begin{cases}
    argmax_{a_t}[Q(s_t,a_t)+H(s_t,a_t)],  & \ if q \leq p \\
    a_{random}, & \ otherwise
    \end{cases}$}
\end{equation}
Where the H-value $H(s_t, a_t)$ influences the choice of action, by evaluating the importance of executing the action $a_t$ (suggested by the heuristic algorithm) having state $s_t$ \cite{DBLP:journals/corr/abs-2106-02757}.
\section{Experiments and Results}

\begin{figure}[!ht]
    \centering
    \includegraphics[width=\columnwidth, height=9cm, trim={0.5cm 10cm 0 3cm},clip]{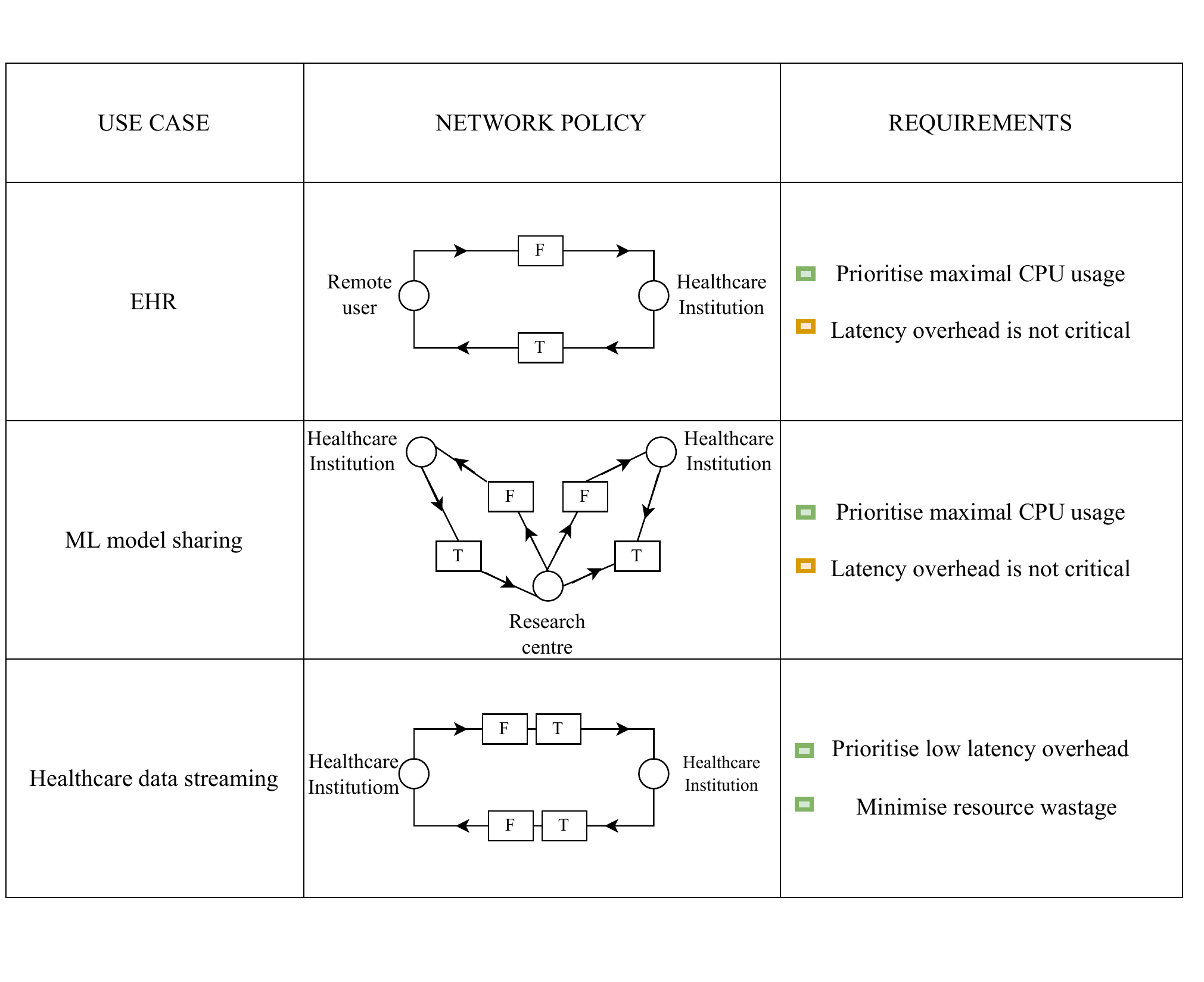}
    \caption{Different network policies and requirements, deploying different use cases; \textbf{F} represents a firewall microservice request, and \textbf{T} represents an encryption microservice request.}
    \label{fig:experiments}
\end{figure}

\textbf{Use cases' Network Policies:}
The security enforcement primitives are consistent throughout all the use cases, such that the \textit{firewall} function is mandated to provide access control to incoming traffic towards the healthcare institution. Likewise, all outgoing traffic should be \textit{encrypted} to protect sensitive data. This is showcased in Fig. \ref{fig:experiments}, where you can see the same logic being applied to all the use cases. 

The EHR use case is profiled to be a small load application with an expected average send rate of 100-200 kB/s. The ML-model sharing use case is defined to be, also, a small load use case with an expected (average) periodic send rate of 100-200 kB/s. Additionally, the streaming use case is profiled to be a large load use case with a 1-3 MB/s send rate. Accordingly, the CPU profiles of different BFCs under different use cases' send rates are recorded in \cite{9973638}, and further used via the heuristic algorithm to make provisioning decisions. 

Moreover, the provisioning algorithm should comply with multiple use case requirements. Maximum CPU usage is prioritised with the EHR and ML model sharing use cases, compared to the streaming use case where low latency overhead is required as well. 

\subsection{Experiments}

To evaluate placement and assignment decisions taken by the different provisioning tools, we run the three use cases according to the defined send rate workloads, and associated CPU profiles. We first reconfigure the graph $G$ as Kubernetes clusters, to reflect Fig. 1 and 4. Connected clusters form a cluster mesh via a cilium backend server \footnote{https://cilium.io/}.  

\textbf{New instances vs CPU utilisation:}
We increase the number of concurrent clients utilising the BFC with locust\footnote{https://locust.io/}, and we record the instantiated new instances/microservice across all the clusters under the different use case configurations, with the growing number of clients. For replication purposes, the implementation and configurations of the experiments are available on GitHub \footnote{https://github.com/epi-project/Netsoft2023}. Furthermore, we compare number of pods (instances) to the actual utilized CPU per instance, to evaluate the CPU resource wastage, and further relate that to the latency.


\textbf{Latency Overhead:}
Similarly, we increase the number of concurrent clients running different use cases, and record the latency of processing one request (sending a request, and receiving a reply back). Low latency can be accomplished by providing high-performance networking, but the goal is to evaluate the overhead latency caused by adding network services in between end nodes, and the effect of different provisioning decisions. We try to minimise the delay when deploying the network function request, and ideally the latency overhead $\approx 0 ms$. We collect the latency overhead by calculating the average of 10 queries, then measuring the effect of the chain addition compared to no functions in between.
\subsection{Results}
\begin{figure}[ht!]
    \centering
    \subfloat[\centering The heuristic algorithm placement]{
          \begin{tikzpicture}
        \node (tiger) [anchor=south west, inner sep=0pt] {\includegraphics[width=0.9\columnwidth, trim={0 0 0 0.72cm},clip, height=4.5cm]{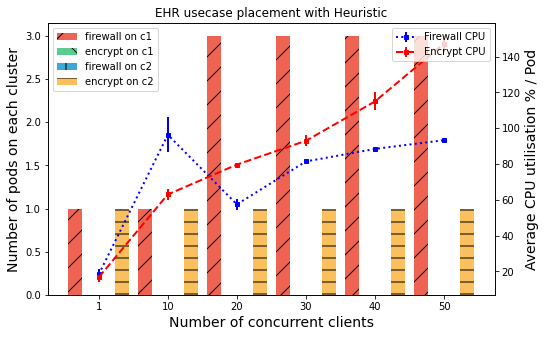} \label{fig:ehr-h} };
        \begin{scope}[x={(tiger.east)},y={(tiger.north west)}]
            \draw [black, dotted] (0.243,0.6) -- (0.243,0);
            \draw [black, dotted] (0.367,0.749) -- (0.368,-0.061);
            \draw [black, dotted] (0.492,0.7) -- (0.492,-0.112);
            \draw [black, dotted] (0.618,0.61) -- (0.618,-0.19);
            \draw [black, dotted] (0.744,0.55) -- (0.744,-0.25);

        \end{scope}
      \end{tikzpicture}}%
    \qquad
    \subfloat[\centering The DQL and HDQL algorithm placement]{\begin{tikzpicture}
        \node (tiger) [anchor=south west, inner sep=0pt] {\includegraphics[width=0.9\columnwidth, trim={0 0 0 0.72cm},clip, height=4.5cm]{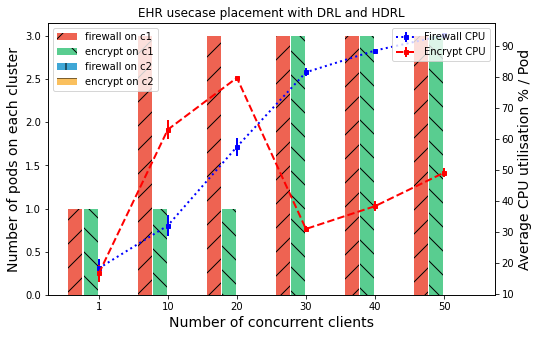} \label{fig:ehr-dql} };
        \begin{scope}[x={(tiger.east)},y={(tiger.north west)}]
            \draw [black, dotted] (0.243,0.6) -- (0.243,0);
            \draw [black, dotted] (0.367,0.749) -- (0.368,-0.061);
            \draw [black, dotted] (0.492,0.7) -- (0.492,-0.112);
            \draw [black, dotted] (0.618,0.61) -- (0.618,-0.19);
            \draw [black, dotted] (0.744,0.55) -- (0.744,-0.25);

        \end{scope}
      \end{tikzpicture}}%
    \caption{ The placement of microservices and the average CPU utilisation running the EHR use case.}%
    \label{fig:ehr-results}%
\end{figure}
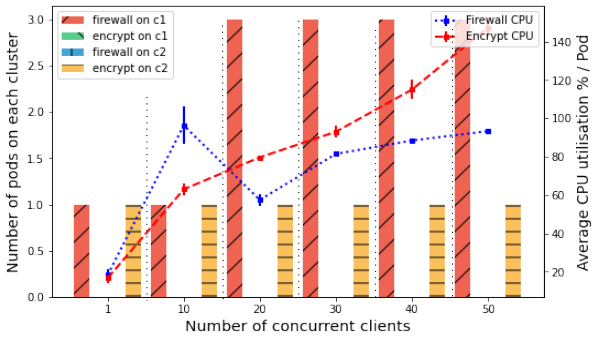
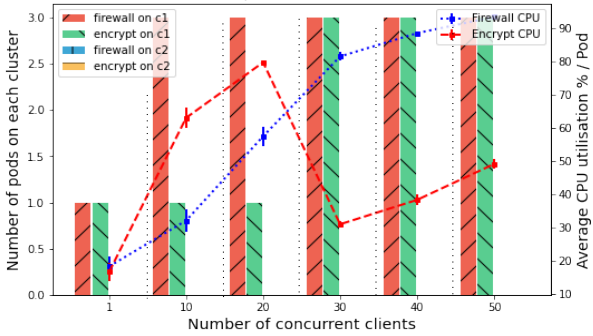

Under the first use case's N-PoPs constraints, no microservice instances are placed on cluster 3 \textit{i.e.} the Research centre cluster (or, if already running, not assigned to the requests running EHR traffic), as shown in Fig. \ref{fig:ehr-results}. The provisioning decision, however, differs with each method, to capture the difference we collect the number of running pods /clusters and the effect of increasing the number of clients from 1 to 50 concurrently running. Moreover, the provisioning decisions as we discussed previously are based on factors: CPU availability, and latency overhead. Hence, we also showcase the average CPU utilisation percentage/pod with each iteration, and the latency recorded to successfully resolve one request.
Firstly, in Fig. \ref{fig:ehr-results}(a) we notice that the heuristic-based placement starts by placing small configured firewall and encryption microservices on cluster 1 and 2, respectively, with the CPU utilisation average of 20-25\% running 1 client on both. The firewall utilisation ramps up faster than encryption to reach 100\% with 10 concurrent clients, then the heuristic placement reacts to the maximal utilisation by upgrading the configuration of the microservice to reach 3 pods running on cluster 1 with 20 concurrent clients running. The trend we notice throughout this plot is that the heuristic placement is more concerned with maximising the average CPU utilisation, such that we end up overcommitting the firewall microservice and assigning it to the maximal number of traffic. The effect of overcommitting the firewall microservices (120\%, 140\% with 40 and 50 clients) is also reflected in the latency in Fig. \ref{fig:lat-results}(a), such that latency overhead increases from 2.21 ms with 30 clients to 5.32 and 6.26ms with 40 and 50 clients, respectively.

On the other hand, with the DQL approach, this latency increase is avoided because the resources are never overcommitted, with the hyperparameters in Eq. \ref{eq:reward} set so that $\alpha > \beta$, so the model will prioritise resources. With this approach, two main decisions to optimise placement are taken, the first is to place the encryption microservice on cluster 1, and by that only using the Kubernetes backend discovery service once, such that microservices on the same cluster belong to the same private network. Another thing is that the DQL is more adaptive to resource bursts, with no need for profiling, and relying on accurate profiles of CPU usage. This is also reflected in the latency, such that the overhead is almost halved compared to the heuristic approach, and it is from 5.32 to 1.26ms running 40 clients. 

With the HDQL approach, there are no major differences, and we end up with the same latency as well. The reason is that the hyperparameters are set in the action policy. With this use case, latency overhead is not crucial, and hence occasional under-provisioning of resources is not an issue, and the heuristic placement is a sufficient tool for this use case.

\begin{figure}[ht!]
    \centering
    \subfloat[\centering The heuristic algorithm placement]{\begin{tikzpicture}
        \node (tiger) [anchor=south west, inner sep=0pt] {{\includegraphics[width=0.9\columnwidth, trim={0 0 0 0.72cm},clip, height=4.5cm]{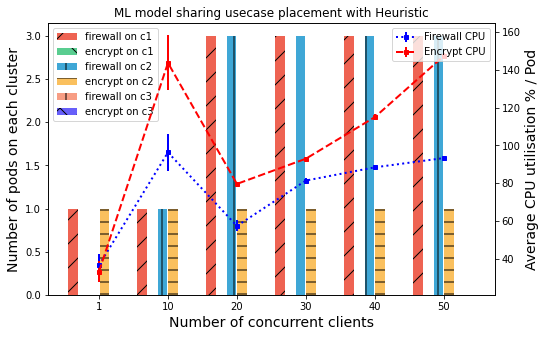} \label{fig:ml-h} } };
        \begin{scope}[x={(tiger.east)},y={(tiger.north west)}]
            \draw [black, dotted] (0.223,0.52) -- (0.223,0);
            \draw [black, dotted] (0.346,0.749) -- (0.346,-0.056);
            \draw [black, dotted] (0.468,0.7) -- (0.468,-0.112);
            \draw [black, dotted] (0.592,0.62) -- (0.592,-0.182);
            \draw [black, dotted] (0.718,0.57) -- (0.718,-0.25);

        \end{scope}
      \end{tikzpicture}
    }%
    \qquad
    \subfloat[\centering The DQL and HDQL algorithm placement ]{\begin{tikzpicture}
        \node (tiger) [anchor=south west, inner sep=0pt] {{\includegraphics[width=0.9\columnwidth, trim={0 0 0 0.72cm},clip, height=4.5cm]{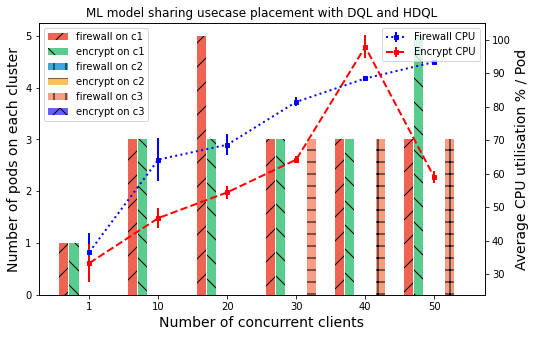} \label{fig:ml-dql} } };
        \begin{scope}[x={(tiger.east)},y={(tiger.north west)}]
            \draw [black, dotted] (0.223,0.52) -- (0.223,0);
            \draw [black, dotted] (0.346,0.749) -- (0.346,-0.056);
            \draw [black, dotted] (0.468,0.7) -- (0.468,-0.112);
            \draw [black, dotted] (0.592,0.62) -- (0.592,-0.182);
            \draw [black, dotted] (0.718,0.57) -- (0.718,-0.25);

        \end{scope}
      \end{tikzpicture}}%
        \qquad
    \caption{ The placement of microservices and the average CPU utilisation running the ML model sharing use case.}%
    \label{fig:ml_stream}%
\end{figure}
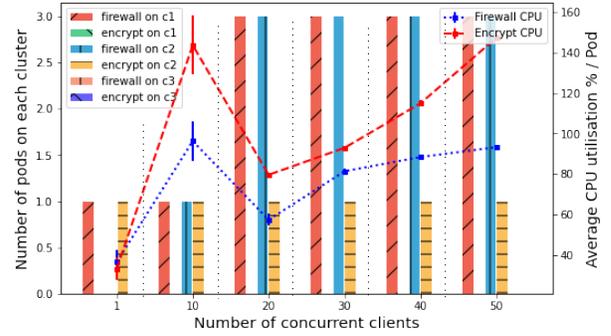
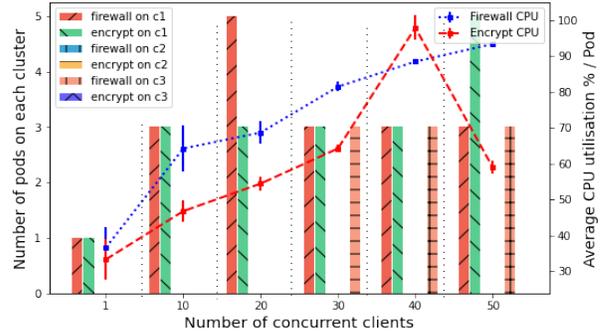
\begin{figure}[ht!]
    \centering
    \subfloat[\centering The heuristic algorithm placement]{
    \begin{tikzpicture}
        \node (tiger) [anchor=south west, inner sep=0pt] {{\includegraphics[width=0.9\columnwidth, trim={0 0 0 0.72cm},clip, height=4.5cm]{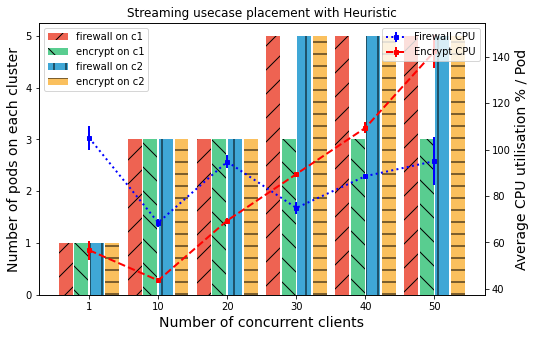} \label{fig:stream-h} } };
        \begin{scope}[x={(tiger.east)},y={(tiger.north west)}]
            \draw [black, dotted] (0.226,0.61) -- (0.226,0);
            \draw [black, dotted] (0.354,0.749) -- (0.354,-0.061);
            \draw [black, dotted] (0.478,0.67) -- (0.478,-0.112);
            \draw [black, dotted] (0.606,0.61) -- (0.606,-0.19);
            \draw [black, dotted] (0.73,0.55) -- (0.73,-0.25);

        \end{scope}
      \end{tikzpicture}}%
    \qquad
    \subfloat[\centering The DQL algorithm placement ]{    \begin{tikzpicture}
        \node (tiger) [anchor=south west, inner sep=0pt] {{\includegraphics[width=0.9\columnwidth, trim={0 0 0 0.72cm},clip, height=4.5cm]{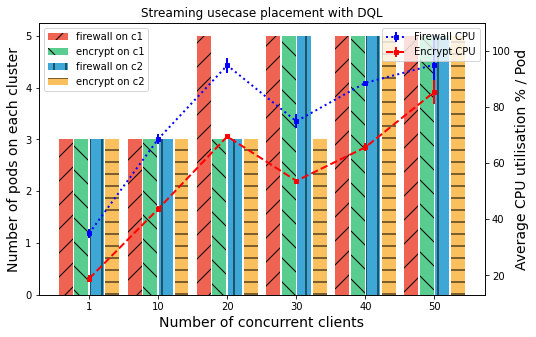} \label{fig:stream-dql} } };
        \begin{scope}[x={(tiger.east)},y={(tiger.north west)}]
            \draw [black, dotted] (0.226,0.61) -- (0.226,0);
            \draw [black, dotted] (0.354,0.749) -- (0.354,-0.061);
            \draw [black, dotted] (0.478,0.67) -- (0.478,-0.112);
            \draw [black, dotted] (0.606,0.61) -- (0.606,-0.19);
            \draw [black, dotted] (0.73,0.55) -- (0.73,-0.25);

        \end{scope}
      \end{tikzpicture}}%
        \qquad
    \subfloat[\centering The HDQL algorithm placement]{    \begin{tikzpicture}
        \node (tiger) [anchor=south west, inner sep=0pt] {{\includegraphics[width=0.9\columnwidth, trim={0 0 0 0.72cm},clip, height=4.5cm]{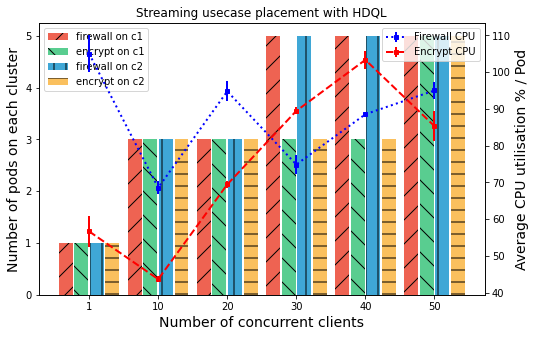} \label{fig:stream-hdql} } };
        \begin{scope}[x={(tiger.east)},y={(tiger.north west)}]
            \draw [black, dotted] (0.226,0.61) -- (0.226,0);
            \draw [black, dotted] (0.354,0.749) -- (0.354,-0.061);
            \draw [black, dotted] (0.478,0.67) -- (0.478,-0.112);
            \draw [black, dotted] (0.606,0.61) -- (0.606,-0.19);
            \draw [black, dotted] (0.73,0.55) -- (0.73,-0.25);

        \end{scope}
      \end{tikzpicture}}%
    \caption{The placement of microservices and the average CPU utilisation running the streaming use case.}%
    \label{fig:stream-results}%
\end{figure}
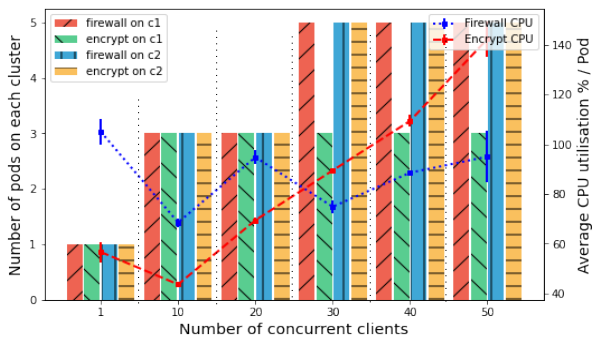
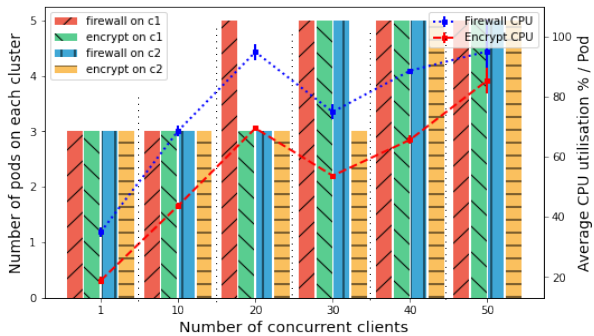
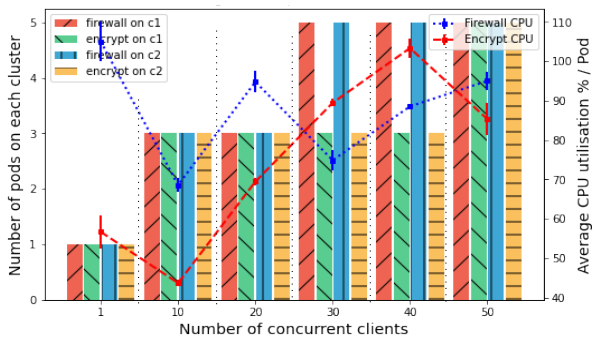

\begin{figure}[ht]
    \centering
    \subfloat[\centering The EHR use case. ]{{\includegraphics[width=0.9\columnwidth, height=5cm]{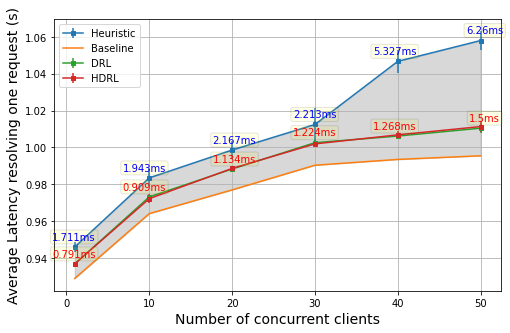} \label{fig:lat-ehr} }}%
    \qquad
    \subfloat[\centering The ML model sharing use case.]{{\includegraphics[width=0.9\columnwidth, trim={0 0 0 0.72cm},clip, height=4.5cm]{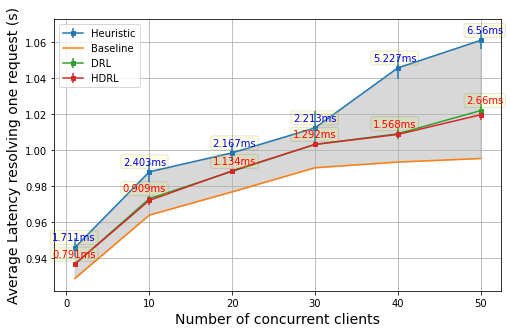} \label{fig:lat-drl} }}%
        \qquad
    \subfloat[\centering The streaming use case]{{\includegraphics[width=0.9\columnwidth, height=4.5cm]{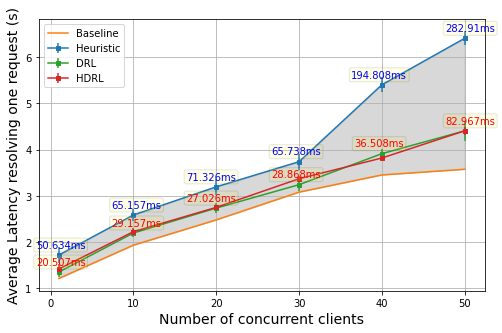} \label{fig:lat-hdrl} }}%
    \caption{ The latency average recorded with different placement methods, and the overhead compared to proxying traffic without passing through extra microservices.}%
    \label{fig:lat-results}%
\end{figure}
Similarly, the ML model sharing use case has initially a similar placement, the difference is that the utilisation is doubled indicating that the pods are assigned to handle double the requests. After that, we start another instance of a firewall on cluster 2, to handle half of the traffic. The heuristic function upgrades the placement configuration to adapt to the increasing workload. Similarly, this approach ends up overcommitting resources and the overhead increases as shown in Fig. \ref{fig:lat-results}(b).

The optimisation decisions taken by the DQL and HDQL approaches are demonstrated by deciding to place microservices on the same cluster, to then start new instances of the firewall cluster 3, that has currently associated with lower round-trip latency, compared to cluster 2. This approach adapts to resource usage bursts more accurately and minimises latency overhead accordingly.

With the last use case, the healthcare data streaming use case, we prioritise latency, and there are three different placement decisions taken. First, the heuristic performs poorly with this use case as shown in Fig. \ref{fig:lat-results}(c), where the chaining is still distributed, and assignment decisions require two different lookups. Unlike the other two approaches, the placement is distributed across clusters, but the chaining is on one cluster. With that, Kubernetes backend microservice discovery is optimised. With this experiment, due to the hyperparameters change, we end up wasting resources with the DQL approach as shown in Fig. \ref{fig:stream-results}(b). Although with these provisioning decisions, we provide the best latency overhead, there exist better actions and hence we have a third different placement with HDQL. The HDQL provisioning provides approximately equal latency with maximal CPU utilisation (shown in Fig. \ref{fig:stream-results}(c) and \ref{fig:lat-results}(c)).


As a result, the heuristic-based approach proved to be sufficient with the EHR and ML model-sharing use cases, where we end up under-provisioning but inflicting tolerable latency. That is especially true since low latency overhead is not crucial in running said use cases, instead, we prioritise minimal CPU wastage. The HDQL tool performs the best while running the streaming use case, where minimal overhead latency was achieved, but with seemingly no over-provisioning and resource wastage. 
\section{Related work}
SFC and VNF provisioning problems have been formalised and addressed by employing heuristics or DQL in the past. Recently, \cite{9449831} formalised TO-DG heuristic-based approach to maximise the network throughput, while considering resource overhead. They consider the required CPU consumption of a VNF, link capacity, and maximum tolerable delay to search for optimal provisioning decisions. 
\cite{9657124} takes a different approach and utilised DQL neural networks to outperform linear programming approaches in an effort to solve the SFC resource allocation problem. Similarly, the authors in \cite{https://doi.org/10.48550/arxiv.1908.03242} use RL-based techniques to formalise the same problem as an MDP and address it with a policy gradient learning agent. We add to the current literature three main contributions: 1) We specifically reason about the type of SFC that is being provisioned to enforce network security policy in DHT use cases, 2) We propose heuristic-boosted DQL techniques to guide and facilitate the learning process according to prior knowledge of profiled data, 3) We add dynamic constraints to the provisioning problem; dynamic N-PoP candidate and prioritising different metrics with different use cases.
\section{Conclusion}
To run DHT use cases, it is essential to first consider data-sharing policies (including network policies), and translate them into actionable service function chain requests; which we call BFC. Additionally, the provisioning of BFC (new instance placement and/or assignment of an incoming request to an old running instance) depends on the use case's requirements and the current state of the infrastructure. We need to prioritise latency or/and minimising resource wastage, and we do that by modelling the decision as a constrained optimising problem. Initially, we address the requirements and constraints via heuristic, which can query the infrastructure state and output decisions influences by the BFC CPU profiles. Next, we propose using DQL methods, which provides more resilience to un-profiled bursts and network degradation. Finally, we combine both approaches to introduce a "best of both world" solution, that proved to be most valuable to accomplish lower latency overhead, with minimal CPU wastage.

The provisioning tools should consider a combination of constraints and objectives, as demonstrated in Section \ref{provision_model}. The framework we propose can be used within any general context, and it effectively provisions network resources to deploy DHT use cases. We conclude that heuristic-based approaches are sufficient when the latency overhead is not crucial, while HDQL tools are most effective otherwise. In future work, we aim to increase the complexity of the service requests; we consider line path SFC in the experiments, but we aim to include: i) requests with bifurcated path with different end-points ii) bifurcated path with single end-point. We also plan to deploy and experiment on test beds with larger topologies and evaluate the orchestrators within larger networks. 

\bibliographystyle{IEEEbib}
\bibliography{strings}

\begin{thebibliography}{10}

\bibitem{gelernter_1991}
David~Hillel Gelernter,
\newblock {\em Mirror World: Or the day software puts in the universe in a
  shoebox ...: How it will happen and what it will mean},
\newblock Oxford Univ. Press, 1991.

\bibitem{grieves_2006}
M.~Grieves,
\newblock {\em Product Lifecycle Management:driving the next generation of Lean
  Thinking},
\newblock McGraw-Hill, 2006.

\bibitem{grieves_2005}
Michael~W. Grieves,
\newblock ``Product lifecycle management: The new paradigm for enterprises,''
\newblock {\em International Journal of Product Development}, vol. 2, no. 1/2,
  pp. 71, 2005.

\bibitem{kamel}
Maged~N. Kamel~Boulos and Peng Zhang,
\newblock ``Digital twins: From personalised medicine to precision public
  health,''
\newblock {\em Journal of Personalized Medicine}, vol. 11, no. 8, pp. 745,
  2021.

\bibitem{mashaly}
Maggie Mashaly,
\newblock ``Connecting the twins: A review on digital twin technology \&amp;
  its networking requirements,''
\newblock {\em Procedia Computer Science}, vol. 184, pp. 299–305, 2021.

\bibitem{9211394}
Jamila~Alsayed Kassem, Cees De~Laat, Arie Taal, and Paola Grosso,
\newblock ``The epi framework: A dynamic data sharing framework for healthcare
  use cases,''
\newblock {\em IEEE Access}, vol. 8, pp. 179909--179920, 2020.

\bibitem{laborde_kamel_barre_benzekri_2007}
Romain Laborde, Michel Kamel, François Barrère, and Abdelmalek Benzekri,
\newblock ``Implementation of a formal security policy refinement process in
  wbem architecture,''
\newblock {\em Journal of Network and Systems Management}, vol. 15, no. 2, pp.
  241–266, 2007.

\bibitem{9973688}
Christopher~A. Esterhuyse, Tim Müller, L.~Thomas Van~Binsbergen, and Adam
  S.~Z. Belloum,
\newblock ``Exploring the enforcement of private, dynamic policies on medical
  workflow execution,''
\newblock in {\em 2022 IEEE 18th International Conference on e-Science
  (e-Science)}, 2022, pp. 481--486.

\bibitem{DBLP:journals/corr/abs-2106-02757}
Ching{-}An Cheng, Andrey Kolobov, and Adith Swaminathan,
\newblock ``Heuristic-guided reinforcement learning,''
\newblock {\em CoRR}, vol. abs/2106.02757, 2021.

\bibitem{9973638}
Jamila~Alsayed Kassem, Adam Belloum, Tim Müller, and Paola Grosso,
\newblock ``Utilisation profiles of bridging function chain for healthcare use
  cases,''
\newblock in {\em 2022 IEEE 18th International Conference on e-Science
  (e-Science)}, 2022, pp. 475--480.

\bibitem{9449831}
Yi~Yue, Bo~Cheng, Meng Wang, Biyi Li, Xuan Liu, and Junliang Chen,
\newblock ``Throughput optimization and delay guarantee vnf placement for
  mapping sfc requests in nfv-enabled networks,''
\newblock {\em IEEE Transactions on Network and Service Management}, vol. 18,
  no. 4, pp. 4247--4262, 2021.

\bibitem{9657124}
Tom~Jenno Wassing, Danny De~Vleeschauwer, and Chrysa Papagianni,
\newblock ``A machine learning approach for service function chain embedding in
  cloud datacenter networks,''
\newblock in {\em 2021 IEEE 10th International Conference on Cloud Networking
  (CloudNet)}, 2021, pp. 26--32.

\bibitem{https://doi.org/10.48550/arxiv.1908.03242}
Jaehoon Koo, Veena~B. Mendiratta, Muntasir~Raihan Rahman, and Anwar Walid,
\newblock ``Deep reinforcement learning for network slicing with heterogeneous
  resource requirements and time varying traffic dynamics,'' 2019.

\end{thebibliography}

\end{document}